\documentclass{elsart}
\usepackage{algorithm,algorithmic,amsmath,amssymb,framed,graphicx,multirow}
\usepackage{graphicx,amssymb}
\usepackage{algorithmic}
\usepackage{algorithm}

\usepackage{longtable}
\usepackage{booktabs}

\usepackage[switch,pagewise]{lineno}
\usepackage[usenames]{color}
\usepackage{url}

\begin{document}
\begin{frontmatter}

\title{A memetic algorithm for the minimum sum coloring problem}
\author{Yan Jin},
\author{Jin-Kao Hao\corauthref{cor}},
\corauth[cor]{Corresponding author.} \ead{hao@info.univ-angers.fr}
\author{Jean-Philippe Hamiez}

\address {LERIA, Universit\'{e} d'Angers, 2 Boulevard Lavoisier, 49045 Angers, France}

\date{March 13, 2013}

\maketitle

\begin{abstract}
Given an undirected graph $G$, the Minimum Sum Coloring problem (MSCP) is to find a legal assignment of colors (represented by natural numbers) to each vertex of $G$ such that the total sum of the colors assigned to the vertices is minimized. This paper presents a memetic algorithm for MSCP based on a tabu search procedure with two neighborhoods and a multi-parent crossover operator. Experiments on a set of 77 well-known DIMACS and COLOR 2002-2004 benchmark instances show that the proposed algorithm achieves highly competitive results in comparison with five state-of-the-art algorithms. In particular, the proposed algorithm can improve the best known results for 17 instances. We also provide upper bounds for 18 additional instances for the first time.

\emph{Keywords}: Sum coloring, memetic algorithm, heuristics, combinatorial optimization
\end{abstract}

\end{frontmatter}

\section{Introduction}
\label{Sec_Intro}


Let $G = (V, E)$ be a simple undirected graph (without loops) with vertex set $V = \{v_1, \ldots , v_n\}$ and edge set $E \subset V \times V$. A proper $k$-coloring $c$ of $G$ is a mapping $c: V \to \{1, \ldots, k\}$ such that $c(v_i) \neq c(v_j)~\forall \{v_i, v_j\} \in E$. A legal or proper $k$-coloring can also be defined as a partition of $V$ into $k$ independent sets or stables $V_1, \ldots, V_k$ such that $\forall u,v \in V_i$ $(i=1, \ldots, k), \{u,v\} \notin E$. The classical graph coloring problem (GCP) aims at finding a proper $k$-coloring with $k$ minimum. 

This paper is dedicated to the NP-hard Minimum Sum Coloring Problem (MSCP) \cite{KubickaPhD1989,Kubicka&Schwenk1989}, which is closely related to GCP. The objective of MSCP is to find a proper $k$-coloring which minimizes the sum of the colors assigned to the vertices. This minimum is the chromatic sum $\Sigma(G)$ of $G$: $\Sigma(G)=\min_{c \in \mathcal{C}} f(c)$, with $\mathcal{C}$ the set of all proper $k$-colorings of $G$ (for all possible $k$ values) and $f(c) = \sum_{i=1}^n c(v_i)$, or $f(c) = \sum_{l=1}^k l|V_l|$ equivalently (where $|V_l|$ is the cardinality of $V_l$), the ``coloring sum'' of the proper $k$-coloring $c$. MSCP applications include VLSI design, scheduling, and resource allocation \cite{Malafiejski2004}.

Considering the theoretical intractability of MSCP, a number of heuristic algorithms have been proposed to find suboptimal solutions, such as a parallel genetic algorithm \cite{Kokosinski&Kwarciany2007}, two greedy heuristics \cite{Li&al2009}, a tabu search metaheuristic \cite{Bouziri&Jouini2010}, a hybrid algorithm (HA) \cite{Douiri&Elbernoussi2011}, an effective heuristic algorithm (EXSCOL) \cite{Wu&Hao2012}, a local search heuristic (MDS5) \cite{Helmar&Chiarandini2011}, and a breakout local search (BLS) \cite{Benlic&Hao2012}. To our knowledge, EXSCOL, HA, MDS5, and BLS are the state-of-the-art algorithms in the literature. EXSCOL is based on extracting large disjoint independent sets and is particular effective for handling large graphs (with at least 500 vertices). HA is a hybrid algorithm which combines a genetic algorithm with a surrogate constraint heuristic. MDS5 is based on variable neighborhood search and iterated local search. BLS combines local search with adaptive perturbation mechanisms to ensure the quality of solutions. 

This paper introduces a Memetic Algorithm for the minimum Sum Coloring problem (MASC), which relies on three key components. First, a double-neighborhood tabu search procedure is especially designed for MSCP (DNTS). DNTS is based on a token-ring application of two complementary neighborhoods to explore the search space and a perturbation strategy to escape from local optima. Second, a multi-parent crossover operator is used for solution recombination. Basically, it tries to transmit large color classes from the parents to the offspring. Finally, a population updating mechanism is devised to determine how the offspring solution is inserted into the population.

We evaluate the performance of MASC on 77 well-known graphs from DIMACS and COLOR 2002-2004 graph coloring competitions (59 of them have been used previously for evaluating sum coloring algorithms). The computational results show that MASC can frequently match the best known results in the literature for most of the 59 cases. In particular, it improves the previous best solution for 17 graphs for which an upper bound is known. For the additional 18 new graphs, we report computational results for the first time.

The paper is organized as follows. Next section describes the general framework and the components of our MASC memetic algorithm, including the population initialization,  the crossover operator and the double-neighborhood tabu search procedure. Detailed computational results and comparisons with five state-of-the-art algorithms are presented in Section \ref{Sec_Results}. Before concluding, Section \ref{Sec_analysis_MASC} investigates and analyzes two key issues of the proposed memetic algorithm.

\section{MASC: A Memetic Algorithm for Minimum Sum Coloring}
\label{Sec_Approach}

A memetic algorithm is a population-based approach where the traditional mutation operator is replaced by a local search procedure \cite{Moscato&Cotta2003,Neri&al2012}. Memetic algorithms are among the most powerful paradigms for solving NP-hard combinatorial optimization problems. In particular, they have been successfully applied to the tightly related GCP \cite{Galinier&Hao1999,Lu&Hao2010,Malaguti&al2008,Porumbel&al2010}.

Our MASC algorithm is summarized in Algorithm \ref{Algo_MASC}. After population initialization, MASC repeats a series of generations (limited to $MaxGeneration$) to explore the search space which is defined by the set of all proper $k$-colorings ($k$ is not a fixed value, Section \ref{Search Space}). At each generation, two or more parents are selected at random (line 6) and used by the dedicated crossover operator to generate an offspring solution (line 7, Section \ref{subsec_sol_crossover_operator}). The offspring solution is then improved by a double neighborhood tabu search (line 8, Section \ref{subsec_sol_TS}). If the improved offspring has a better sum of colors, it is then used to update the current best solution found so far (lines 9-10). Finally, the population updating criterion decides whether the improved offspring will replace one existing individual of the population or not (line 12, Section \ref{subsec_sol_population_update}).

\begin{algorithm}
\footnotesize
\caption{An overview of the MASC memetic algorithm for MSCP}\label{Algo_MASC}
\begin{algorithmic}[1]
   \STATE \textbf{input}: A graph $G$
   \STATE \textbf{output}: The minimum sum coloring $c_*$ found and its objectif $f(c_*)$ 
   \STATE Population\_Initialization($P,p$) /* Population $P$ has $p$ solutions, Sect. \ref{subsec_sol_initial_population} */
   \STATE $f_* \gets \min_{c \in P} f(c)$ /* $f_*$ records the best objective value found so far */
   \FOR{$i \gets 1$ \textbf{to} $MaxGeneration$}  
       \STATE $P' \gets \textrm{Selection}(P)$ /* Select 2 or more parents at random for crossover */
        \STATE $o \gets \textrm{Crossover}(P')$ /* Crossover to get an offspring solution, Sect. \ref{subsec_sol_crossover_operator} */
        \STATE $o \gets \textrm{DNTS}(o)$
        		/* Improve $o$ with the DNTS procedure, Sect.~\ref{subsec_sol_TS}*/
        \IF{$f(o) < f_*$}
           \STATE $f_* \gets f(o); c_*  \gets o$
        \ENDIF
        \STATE Population\_Updating($P, o$) /* Sect. \ref{subsec_sol_population_update} */ 
   \ENDFOR
   \RETURN $f_*, c_*$
\end{algorithmic}
\end{algorithm}

\subsection{Search Space and Evaluation Function}
\label{Search Space}

The search space explored by MASC is the set $\mathcal{C}$ of all \textsl{proper} $k$-colorings of $G$ ($k$ is not fixed). For a given proper $k$-coloring $c$, its quality is directly assessed by the sum of colors $f(c) = \sum_{v \in V} c(v) = \sum_{l = 1}^k l|V_l|$. 

\subsection{Initial Population}
\label{subsec_sol_initial_population}

Our algorithm begins with a population $P$ of $p$ feasible colorings. This population can be obtained by any graph coloring algorithm that is able to generate different proper colorings for a graph. In our case, we employ the well-known TABUCOL \cite{Hertz&DeWerra1987}, more precisely its improved version introduced in \cite{Galinier&Hao1999}. For a given graph $G$, TABUCOL tries to find a proper $k$-coloring with $k$ as small as possible. Given the stochastic nature of TABUCOL, we run it multiple times to obtain different $k$-colorings ($k$ may take different values). Each resulting $k$-coloring is inserted into $P$ if it is not already present (discarded otherwise). This process is repeated until $P$ is filled with $p$ different $k$-colorings.


\subsection{Crossover Operator}
\label{subsec_sol_crossover_operator}

The crossover operator is an important component in a population-based algorithm. It is used to generate one or more new offspring individuals to discover new promising search areas.

MASC uses a multi-parent crossover operator, called MGPX, which is similar to the one introduced in \cite{Hamiez&Hao2002} as a variant of the well-known GPX crossover first proposed in \cite{Galinier&Hao1999} for GCP (restricted to two parents). MGPX generates only one offspring solution $o$ from $\alpha$ parents randomly chosen from $P$, where $\alpha$ varies from 2 to 4 according to $|V|$ and the best $k$-coloring found for GCP (see Eq. \ref{EqNbParents}). Motivations for these $\alpha$ values can be found in \cite{Porumbel&al2010}.

\begin{equation}
\alpha= \left \{ \begin{array} {ll}
2,   &   \textrm{if $|V| / k < 5$} \\
3,   &    \textrm{if $5 \le |V| / k \le 15$}\\
4,   &    \textrm{otherwise}
\end{array}  \right.
\label{EqNbParents}
\end{equation}

\begin{algorithm}[h]
\footnotesize
\caption{Pseudo-code of the MGPX combination operator}
\label{AlgoMGPX}
\begin{algorithmic}[1]
\STATE \textbf{input}: A set $P'$ of $\alpha$ parents randomly chosen from $P$
\STATE \textbf{output}: An offspring $o$
\STATE $\nu \gets 0$ /* Counts the number of colored vertices in $o$ */
\STATE $\kappa \gets 0$ /* Counts the number of colors used in $o$ */
\STATE $\varphi_j \gets 0~\forall c_j \in P'$ /* To identify which parents are forbidden for color $\kappa$ */
\WHILE{$\nu < |V|$}
	\STATE $\kappa \gets \kappa + 1$
	\STATE $P_\kappa \gets \{c_j \in P': \varphi_j < \kappa\}$ /* Set of parents allowed for color $\kappa$ */
	\STATE Let $V_*^j$ be a color class of maximum cardinality $\forall c_j \in P_\kappa$
	\STATE $\nu \gets \nu + |V_*^j|$
	\FORALL{$v \in V_*^j$}
		\STATE $o(v) \gets \kappa$
		\STATE Remove $v$ from $c_i~\forall c_i \in P'$
	\ENDFOR
	\STATE $\varphi_j \gets \kappa + \lfloor \alpha / 2 \rfloor$ /* Forbid using $c_j \in P'$ for a few steps */
\ENDWHILE
\RETURN $o$
\end{algorithmic}
\end{algorithm}

MGPX is summarized in Algorithm \ref{AlgoMGPX}. It builds the color classes of the $o$ offspring one by one, transmitting as many vertices as possible from the parents at each step (for quality purpose) (lines 7-14). Once a parent has been used for transmitting an entire color class to $o$, the parent is not considered for a few steps with the purpose of varying the origins of the color classes of $o$ (line 15). This strategy avoids transmitting always from the same parent and introduces some diversity in $o$ \cite{Lu&Hao2010}. Note that the offspring solution is always a proper $k$ coloring while the number of colors used by the offspring can be higher than those of the considered parents.




\subsection{A Double-Neighborhood Tabu Search for Sum Coloring}
\label{subsec_sol_TS}

Local optimization is another important element within a memetic algorithm. In our case, its role is to improve as far as possible the quality (i.e., the sum of colors) of a given solution returned by the MGPX crossover operator. This is achieved by a Double-Neighborhood Tabu Search (DNTS) procedure specifically designed for MSCP (see Algorithm \ref{Algo_tabu}). 

DNTS uses two different and complementary neighborhoods $N_1$ and $N_2$ which are applied in a token-ring way \cite{GasperoSchaerf2006,Luetal2011} to find good local optima (intensification) (lines 2-14). More precisely, we start our search with one neighborhood (lines 6-9) and when the search ends with its best local optimum, we switch to the other neighborhood to continue the search while using the last local optimum as the starting point (lines 10-13). When this second search terminates, we switch again to the first neighborhood and so on. DNTS continues the exploration of each neighborhood $N_i$ ($i=1,2$) until $p_i$ ($i=1,2$) consecutive iterations fail to update the best solution found. 

This neighborhood-based intensification phase terminates if the best local optimum is not updated for $p_3$ consecutive iterations (line 14). At this point, we enter into a diversification phase by triggering a perturbation to escape from the local optimum (line 15, Section \ref{subsec_sol_perturbation_mechanism}). The DNTS procedure stops when a maximum number of iterations $p_4$ is met. We explain below the two neighborhoods, the tabu list management and the perturbation mechanism.

\begin{algorithm}[h]
\begin{small}
 \caption{Pseudo-code of double-neighborhood tabu search for MSCP}\label{Algo_tabu}
 \begin{algorithmic}[1]
   \STATE \sf \textbf{Input}: Graph $G$, a $k$-coloring $c$
   \STATE \textbf{Output}: the best improved $k$-coloring
   \STATE $c_* \leftarrow c$  
   \WHILE{a stop condition is not met}
       \REPEAT
            \STATE  $c$ $\leftarrow$ TS($N_1, c$)   \   /* Tabu search with neighborhood $N_1$, Sect. \ref{SectN2} */
           \IF{$f(c)<f(c_* )$} 
                   \STATE $c^* \leftarrow c$                  
          \ENDIF

          \STATE $c$ $\leftarrow$ TS($N_2, c$)     \ \ \ /* Tabu search with neighborhood $N_2$, Sect.  \ref{subsec_sol_one_move_tabu} */
           \IF{$f(c)<f(c_* )$} 
                   \STATE $c_* \leftarrow c$                  
          \ENDIF
     \UNTIL {$c_*$ not improved for $p_3$ consecutive iterations}
     \STATE $c \leftarrow$ Perturbation($c_*$)            \ \ /* Search is stagnating, generate a new starting solution by perturbing the best $k$-coloring found so far, Sect. \ref{subsec_sol_perturbation_mechanism} */
   \ENDWHILE

 \end{algorithmic}
 \end{small}
\end{algorithm}

\subsubsection{$N_1$: A Neighborhood Based on Connected Components}
\label{SectN2}

\begin{figure}[h]
\begin{center}
\begin{tabular}{c@{\qquad\qquad}c}
\includegraphics{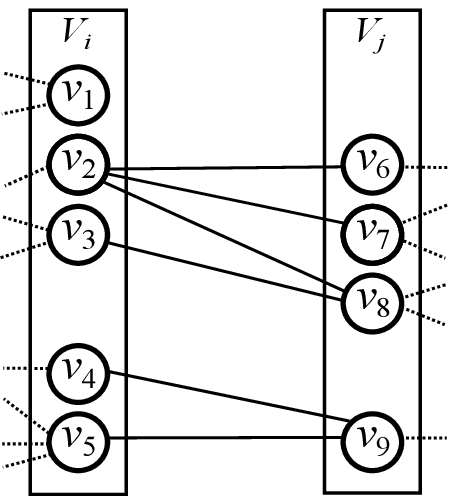} & \includegraphics{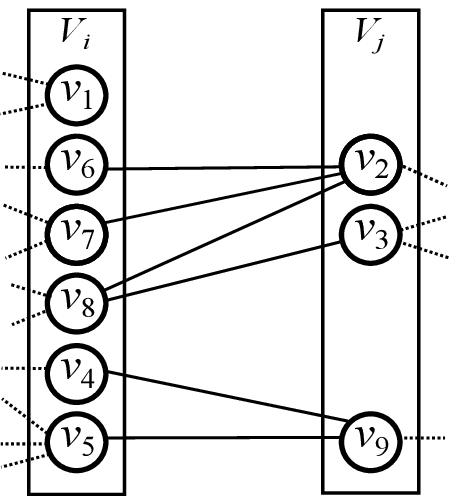}\\
A proper $k$-coloring $c$ & A possible $c' \in N_1(c)$
\end{tabular}
\caption{$N_1$: An illustrative example with two partial colorings ($c$ and $c'$ are restricted here to two $V_i$ and $V_j$ color classes)}
\label{fig_chain_swap}
\end{center}
\end{figure}

The first neighborhood $N_1$ can be described by the operator $Exchange(i,j)$. Given a proper $k$-coloring $c = \{V_1, \ldots, V_k\}$, operator $Exchange(i,j), (1 \le i \neq j \le k)$ swaps some vertices of a color class $V_i$ against some connected vertices of another color class $V_j$. Formally, let $G_{i, j}(c)$ be the set of all connected components of more than one vertex in the subgraph of $G$ induced by color classes $V_i$ and $V_j$ in a proper $k$-coloring $c$. In Figure \ref{fig_chain_swap} (left) for instance, $G_{i, j}(c)$ is composed of two graphs (say $g_1$ and $g_2$): $g_1$ is the subgraph induced by $\{v_2, v_3, v_6, v_7, v_8\}$ and $g_2$ is induced by $\{v_4, v_5, v_9\}$.

Neighborhood $N_1(c)$ is composed of the set $\mathcal{G}(c)$ of all possible elements in all the $G_{i, j}(c)$ sets: $\mathcal{G}(c) = \cup_{1 \le i < j \le k} G_{i, j}(c)$. In other words, $N_1(c)$ includes all the proper $k$-colorings that can be obtained from the current $k$-coloring $c$ by exchanging the vertices of a connected component induced by color classes $V_i$ and $V_j$. Figure \ref{fig_chain_swap} shows an example where by exchanging the two sets of vertices $\{v_2, v_3\}$ and $\{v_6, v_7, v_8\}$ of connected component $g_1$ of the current $k$-coloring $c$ (left drawing), we obtain a neighboring $k$-coloring $c'$ (right drawing).

\subsubsection{$N_2$: A Neighborhood Based on One-Vertex-Move}
\label{subsec_sol_one_move_tabu}

The second neighborhood $N_2$ is conventional and is simpler than $N_1$. $N_2$ can be described by the operator $OneMove(v,i,j)$. Given a proper $k$-coloring $c = \{V_1, \ldots, V_k\}$, operator $OneMove(v,i,j), (1 \le i \neq j \le k)$ displaces one single vertex $v$ of a color class $V_i$ to another color class $V_j$ such that the resulting $k$-coloring remains proper. For instance, from the current coloring $c$ of the left drawing of Figure \ref{fig_chain_swap}, moving vertex $v_1$ from $V_i$ to $V_j$ gives a neighboring solution. Neighborhood $N_2(c)$ is composed of all the possible proper $k$-colorings by applying $OneMove(v,i,j)$ to the current $k$-coloring $c$. Like neighborhood $N_1$, the solutions of this second neighborhood are also proper $k$-colorings. Moreover, the number of colors of the neighboring solutions remains the same as that of the current coloring.


\subsubsection{Neighborhood Examination and Tabu List}
\label{Neighborhood examination}

DNTS applies these two neighborhoods $N_1$ and $N_2$ in a token-ring way \cite{GasperoSchaerf2006,Luetal2011}. The alternation between $N_1$ and $N_2$ is triggered when the current neighborhood is exhausted, i.e., when the current best solution cannot be further improved for a fixed number of consecutive iterations.  

During the exploration of a neighborhood, DNTS uses a tabu list \cite{{Glover&Laguna1997}} to avoid short-term cycles. Precisely, for the neighborhood $N_1$ defined by the operator $Exchange(i,j)$, when a set of vertices of a color class $V_i$ are exchanged with a set of vertices of another color class $V_j$, exchanges between $V_i$ and $V_j$ are forbidden for the next $TT$ iterations (called tabu tenure). For the neighborhood $N_2$ defined by the operator $OneMove(v,i,j)$, when a vertex $v$ of a color class $V_i$ is displaced to another color class $V_j$, the vertex $v$ is forbidden to go back to $V_i$ for the next $TT$ iterations. 

In both cases, the tabu tenure $TT$ is determined simply by taking a random number from $\{0, \ldots k - 1\}$. Moreover, a forbidden $Exchange$ or $OneMove$ operation is always accepted if it leads to a neighboring solution better than the best solution found so far (this is called aspiration according to the tabu search terminology). 

As shown in Algorithm \ref{Algo_tabu}, at each iteration of our DNTS, a best neighboring solution is selected among all the allowed solutions (from $N_1$ or $N_2$) to replace the current solution. The tabu list is updated accordingly after each iteration. 


\subsubsection{The Perturbation Mechanism}
\label{subsec_sol_perturbation_mechanism}

In addition to the basic diversification mechanism of the tabu list, our DNTS algorithm applies a stronger diversification strategy based on perturbations to escape deep local optima. The perturbation is triggered when the current intensification phase cannot update the recorded best solution $c_*$ for $p_3$ consecutive iterations (see line 15, Algorithm \ref{Algo_tabu}). In this case, the search is considered to be trapped in a deep local optimum and a strong diversification is needed to bring the search to a new search region. To achieve this, we apply the following perturbation technique to modify the recorded best solution $c_*$ and then use this perturbed solution to initialize DNTS. Suppose $c_*$ is composed of $k$ different color classes and let $V_l$ be the largest color class. We introduce an additional color class  $V_{k + 1}$ and then move randomly one third of the vertices of $V_l$ into $V_{k + 1}$. In order to prevent the subsequent search from coming back to $c_*$, $V_l$ and $V_{k + 1}$ are classified tabu and cannot take part of an $Exchange$ or a $OneMove$ operation for the next $TT$ iterations (see Section \ref{Neighborhood examination}).


\subsection{Population Updating}
\label{subsec_sol_population_update}

The management of the population usually controls and balances two important factors in population-based heuristics: Quality and diversity.

Quality can naturally be measured here using the coloring sum function ($f$). The proper $k$-coloring $c_i$ is better than $c_j$ if $f(c_i) < f(c_j)$.

We use the following distance $H$ to estimate the diversity. Given two coloring $c_i$ and $c_j$, $H_{i, j}$ is the number of vertices in $c_i$ and $c_j$ which have different colors: $H_{i, j} = |\{v \in V: c_i(v) \neq c_j(v)\}|$. A small $H_{i, j}$ value indicates a high similarity between $c_i$ and $c_j$. $H$ is also employed to measure how much diversity $H_{i, P}$ a particular $k$-coloring $c_i$ contributes to the entire population $P$: $H_{i, P} = \min_{j \neq i} H_{i, j}$. Again, a small (large) $H_{i, P}$ value indicates that $c_i$ adds a low (high) diversity to $P$.

In MASC, $f$ and $H$ are combined in a $s$ ``score'' function which is used to decide whether an offspring solution $o$ replaces an individual in the population $P$ or not: $s(c_i)=f(c_i)+e^{0.08|V|/H_{i,P}}~\forall c_i \in P$. Precisely we first add $o$ into $P$ and compute all $s(c_i)$. We then identify the worse configuration $c_w$ (i.e., $s(c_w)$ is maximum). The replacement strategy applies the following rules:

\begin{description}
\item[Case 1] ($c_w$ is not $o$): Remove $c_w$ from $P$;
\item[Case 2] ($c_w$ is $o$): Remove the second worse individual from $P$ with probability 0.2, and discard $o$ otherwise.
\end{description}

\section{Experimental Results}
\label{Sec_Results}

Our MASC approach was tested on a benchmark composed of 77 well-known graphs commonly used to report computational results for MSCP: 39 are part of the COLOR 2002--2004 competitions and the 38 others are known as ``DIMACS'' instances. Most of these graphs are available on-line from \url{http://mat.gsia.cmu.edu/COLOR04}, except the 6 ``flat'' instances that can be retrieved from \url{http://mat.gsia.cmu.edu/COLOR/instances.html}. The main characteristics of each graph appear in Tables \ref{table_MASC} and \ref{table_MASC_large}, see columns 1--4 (COLOR 2002--2004 instances are at the top of Table \ref{table_MASC} and DIMACS instances at the bottom): Name of the graph, order ($|V|$), size ($|E|$), and chromatic sum $\Sigma$ (or the best known upper bound when $\Sigma$ is unknown).

MASC is programmed in C++ and compiled using GNU gcc on a PC with 2.7 GHz CPU and 4 Gb RAM. Like many memetic algorithms, we use a small population of 10 individuals. The values of the other parameters were determined empirically, see Table \ref{Parameter_Settings}. Notice that $MaxGenerations = 50$ is the stop condition that determines the running time of the algorithm. The best results of our MASC algorithm will be made available at http://www.info.univ-angers.fr/pub/hao/masumcol.html.


\begin{table}[h]
\begin{scriptsize}
\caption{Settings of parameters}
\label{Parameter_Settings}
\begin{tabular}{cclr}
\hline
Parameter & Sect. & Description & Value\\
\hline
$p_1$ &    \ref{subsec_sol_TS}                        & Maximum number of non-improving moves for TS using $N_1$          & 500\\
$p_2$     & \ref{subsec_sol_TS}                        &  Maximum number of non-improving moves for TS using $N_2$& 1\,000              \\
$p_3$     & \ref{subsec_sol_TS}                        & Maximum number of non-improving moves of TS for perturbation & 4\,000  \\
$p_4$     & \ref{subsec_sol_TS}                        & Maximum iterations of TS procedure & 10\,000  \\
$MaxGenerations$ & \ref{subsec_sol_population_update}      & Maximum number of generations                & 50                    \\
\hline
\end{tabular}
\end{scriptsize}
\end{table}

\subsection{Computational Results}
\label{subsec_computational_results}

Columns 6--10 in Table \ref{table_MASC} present detailed computational results of our MASC algorithm: Best result obtained ($\Sigma_*$) with the number of required colors ($k_*$), success rate (SR, percentage of runs such that $\Sigma_* \le \Sigma$), average coloring sum (Avg.), standard deviation ($\sigma$), and average running time to reach $\Sigma_*$ ($t$, in minutes). Column $k$ shows the chromatic number or its best upper bound (i.e., the smallest number of colors for which a $k$-coloring is ever reported).  Furthermore, a sign ``--'' in column 4 ($\Sigma$) indicates that the instance was never tested in literature for the sum coloring problem (there are 18 such cases). The reported values are based on 30 independent runs (i.e., with different random seeds).

\renewcommand{\baselinestretch}{0.7}\large\normalsize
\begin{table}
\begin{scriptsize}
\caption{Detailed computational results of MASC on the set of 39 COLOR 2002-2004 instances (upper part) and 24 DIMACS instances (bottom part)}
\label{table_MASC}
\begin{tabular}{lrrrcrrrrrr}
\hline
\multicolumn{4}{c}{Characteristics of the graphs} && \multicolumn{6}{c}{MASC}\\
\cline{1-4}\cline{6-11}
Name & $|V|$ & $|E|$ & $\Sigma$ && $k$ & $\Sigma_*(k_*)$ &SR& Avg. & $\sigma$ & $t$ \\
\hline
myciel3    & 11 & 20   & 21 &&4& 21(4)  &1.0& 21.0&0.0 & 0.0  \\
myciel4    & 23 & 71  & 45  &&5& 45(5)  &1.0& 45.0&0.0 & 0.0   \\
myciel5    & 47 & 236 &  93 &&6&  93(6)  &1.0& 93.0&0.0 & 0.0 \\
myciel6    & 95 & 755  &   189 &&7&   189(7)  &1.0& 189.0&0.0 & 0.1  \\
myciel7    & 191 & 2\,360  & 381  &&8& 381(8)  &1.0& 381.0&0.0 & 1.1\\
anna    & 138 & 986  & 276 &&11& 276(11)  &1.0& 276.0&0.0 &0.1  \\
david    & 87 & 812  & 237 &&11& 237(11)  &1.0& 237.0&0.0 & 0.1\\
huck    & 74 & 602   & 243 &&11& 243(11)  &1.0& 243.0&0.0 & 0.0\\
jean    & 80 & 508  & 217 &&10& 217(10)  &1.0& 217.0&0.0 &0.0 \\
homer    & 561 & 1\,629  & -  &&10 & \textbf{1\,123(12)}  &1.0 & 1\,136.2&5.8  &80.6 \\
queen5.5    & 25 & 160  & 75    &&5& 75(5)  &1.0& 75.0&0.0 & 0.0  \\
queen6.6    & 36 & 290  & 138 &&7& 138(8) &1.0& 138.0&0.0 & 1.1  \\
queen7.7    & 49 & 476 &  196 &&7&  196(7) &1.0& 196.0&0.0 & 0.0 \\
queen8.8    & 64 & 728 & 291 &&9& 291(9) &1.0& 291.0&0.0 & 12.8   \\
queen9.9    & 81 & 2\,112 & -  &&10& \textbf{409(10)} &0.3 & 410.5&1.2  & 1.2   \\
queen8.12    & 96 & 1\,368 & -  &&12& \textbf{624(12)} &1.0& 624.0&0.0 & 0.0   \\
games120    & 120 & 638 & 443  &&9& 443(9) &1.0& 443.0&0.0 & 0.5\\
miles250    & 128 & 387  & 325  &&8& 325(8) &1.0& 325.0&0.0 & 0.4 \\
miles500    & 128 & 1\,170 & $\le 709$ &&20& \textbf{705(20)} &1.0& 705.0&0.0 & 1.0  \\
fpsol2.i.1    & 496 & 11\,654  &3\,403 &&65&3\,403(65)  &1.0& 3\,403.0&0.0 & 8.7  \\
fpsol2.i.2    & 451 & 8\,691  &-  &&30 &\textbf{1\,668(30)}  &1.0& 1\,668.0 &0.0 & 5.7  \\
fpsol2.i.3    & 425 & 8\,688  &-  &&30 &\textbf{1\,636(30)}  &1.0& 1\,636.0 &0.0 & 7.0  \\
mug88\_1    & 88 & 146 &  178  &&4&  178(4)  &1.0& 178.0&0.0 & 0.1  \\
mug88\_25    & 88 & 146  &  178  &&4&  178(4)  &1.0& 178.0&0.0 & 0.2\\
mug100\_1    & 100 & 166   &  202 &&4&  202(4)  &1.0& 202.0&0.0 & 0.2  \\
mug100\_25    & 100 & 166  &  202   &&4&  202(4)  &1.0& 202.0&0.0 & 0.3\\
2-Insertions\_3    & 37 & 72   &  62 &&4&  62(4)  &1.0& 62.0&0.0 & 0.0  \\
3-Insertions\_3    & 56 & 110  &  92 &&4&  92(4)  &1.0& 92.0&0.0 & 0.0  \\

inithx.i.1   & 864 & 18\,707 &-  &&54 &\textbf{3\,676(54)} &1.0 &3\,676.0 &0.0 &7.6\\
inithx.i.2   & 645 & 13\,979 &-  &&31 &\textbf{2\,050(31)} &1.0 &2\,050.0 &0.0 &4.4\\
inithx.i.3   & 621 & 13\,969 &-  &&31 &\textbf{1\,986(31)} &1.0 &1\,986.0 &0.0 &1.8\\

mulsol.i.1   & 197 & 3\,925 &-  &&49 &\textbf{1\,957(49)} &1.0 &1\,957.0 &0.0 &0.1\\
mulsol.i.2   & 188 & 3\,885 &-  &&31 &\textbf{1\,191(31)} &1.0 &1\,191.0 &0.0 &0.2\\
mulsol.i.3   & 184 & 3\,916 &-  &&31 &\textbf{1\,187(31)} &1.0 &1\,187.0 &0.0 &0.2\\
mulsol.i.4   & 185 & 3\,946 &-  &&31&\textbf{1\,189(31)} &1.0 &1\,189.0 &0.0 &0.2\\
mulsol.i.5   & 186 & 3\,973 &-  &&31 &\textbf{1\,160(31)} &1.0 &1\,160.0 &0.0 &0.2\\

zeroin.i.1    & 211 & 4\,100   & -  &&49&  \textbf{1\,822(49)}  &1.0& 1\,822.0&0.0 & 0.2\\
zeroin.i.2    & 211 & 3\,541   &  1\,004 &&30&  1\,004(30)  &1.0& 1\,004.0&0.0 & 0.1\\
zeroin.i.3    & 206 & 3\,540  &  998 &&30&  998(30)  &1.0& 998.0&0.0 & 0.1  \\
\hline
DSJC125.1    & 125 & 736  & 326 &&5& 326(7)  &0.7& 326.6&0.9 & 4.4  \\
DSJC125.5    & 125 & 3\,891   & 1\,012  &&17& 1\,012(18)  &0.1& 1\,020.0&3.9 & 3.5 \\
DSJC125.9    & 125 & 6\,961   & 2\,503 &&44& 2\,503(44)  &0.5& 2\,508.0&5.6 & 1.9 \\
DSJC250.1    & 250 & 3\,218   & 973 &&8& 974(9)  &0.0& 990.5&8.3 & 17.3   \\
DSJC250.5    & 250 & 15\,668  &3\,219   &&28&3\,230(31)  &0.0& 3\,253.7&14.3 & 23.1 \\
DSJC250.9    &250 & 27\,897  &  $\le 8\,286$ &&72&  \textbf{8\,280(74)}  &0.1& 8\,322.7&22.3 & 5.6  \\

DSJC500.1    & 500 & 12\,458   & 2850 &&12& 2\,940(14)  &0.0& 3\,013.4& 28.3 & 50.4  \\
DSJC500.5    & 500 & 62\,624  &10\,910   &&48&11\,101(53)  &0.0& 11\,303.5 &73.9 & 202.5 \\
DSJC500.9    & 500 & 112\,437  &29\,912 &&126&  29\,994(126)  &0.0& 30059.1&31.6  & 90.9  \\

flat300\_20\_0    & 300 & 21\,375 & 3\,150 &&20& 3\,150(20) &1.0& 3\,150.0&0.0 & 0.0 \\
flat300\_26\_0    & 300 & 21\,633 & 3\,966 &&26& 3\,966(26) &1.0& 3\,966.0&0.0 & 0.8 \\
flat300\_28\_0    & 300 & 21\,695 &  $\le 4\,282$ &&28&  \textbf{4\,238(30)} &0.1& 4\,313.4&22.3 & 309.7 \\
le450\_5a    & 450 & 5\,714  &-  &&5&\textbf{1\,350(5)}  &1.0& 1\,350.0 &0.0 & 0.7  \\
le450\_5b    & 450 & 5\,734  &-  &&5&\textbf{1\,350(5)}   &1.0& 1\,350.0 &0.0 & 0.4  \\
le450\_5c    & 450 & 9\,803  &-  &&5&\textbf{1\,350(5)}   &1.0& 1\,350.0 &0.0 & 0.2  \\
le450\_5d    & 450 & 9\,757  &-  &&5&\textbf{1\,350(5)}   &1.0& 1\,350.0 &0.0 & 0.5  \\
le450\_15a    & 450 & 8\,168  &2\,632 &&15&2\,706(19)  &0.0& 2\,742.6&13.8 & 41.3  \\
le450\_15b    & 450 & 8\,169 & 2\,642  &&15& 2\,724(19) &0.0& 2\,756.2&14.8 & 40.3 \\
le450\_15c    & 450 & 16\,680 & $\le 3\,866$ &&15& \textbf{3\,491(16)} &1.0& 3\,491.0&0.0 & 45.3  \\
le450\_15d    & 450 & 16\,750 &  $\le 3\,921$ &&15&  \textbf{3\,506(17)} &1.0& 3\,511.8&3.6 &59.8 \\
le450\_25a    & 450 & 8\,260 &3\,153 &&25&3\,166(27) &0.0& 3\,176.8&4.4 & 39.2 \\
le450\_25b    & 450 & 8\,263 & 3\,366 &&25& 3\,366(26) &0.1& 3\,375.1&3.4 & 40.3 \\
le450\_25c    & 450 & 17\,343 & 4\,515  &&25& 4\,700(31) &0.0& 4\,773.3&25.2 & 75.3  \\
le450\_25d    & 450 & 17\,425 & 4\,544 &&25& 4\,722(29) &0.0& 4\,805.7&27.4 & 63.4  \\
\hline
\end{tabular}
\end{scriptsize}
\end{table}
\renewcommand{\baselinestretch}{1.0}\large\normalsize

From Table \ref{table_MASC}, one observes that for the 25 cases of 39 COLOR 2002--2004 instances with known upper bounds (see top part of the table), MASC improves the best known upper bound for one instance (miles500) and equals the best known results for the other 24 graphs. For the remaining 14 cases, we provide upper bounds for the first time. Furthermore, MASC achieves robust results here since $\textrm{SR} = 1.0$ and $\sigma = 0.0$ for these graphs except two instances (homer and queen9.9). The average running time of MASC ranges from less than one second to about 13 minutes except for the homer instance.

For the set of 24 DIMACS instances (bottom part), the MASC algorithm improves the best known upper bound for 4 graphs (DSJC250.9, flat300\_28\_0, le450\_15c, and le450\_15d) and equals the best known results for 6 instances. Unfortunately, MASC was unable to reach the best known results for the other 10 graphs (see lines where $\textrm{SR} = 0.0$). For the four graphs which were not tested previously, we report upper bounds achieved for the first time. The average running time is less than 76 minutes except for the DSJC500.5 and flat300\_28\_0 instances. Finally, we notice that the number of colors needed to ensure the best sum coloring ($k_*$) can be larger than the chromatic number or its best upper bound ($k$).

\subsection{Comparisons With State-of-the-art Algorithms}
\label{subsec_comparison_results}

Table \ref{table_many_algorithms} compares MASC with 5 recent effective algorithms that cover the best known results for the considered benchmark: EXSCOL \cite{Wu&Hao2012}, BLS \cite{Benlic&Hao2012}, MDS5 \cite{Helmar&Chiarandini2011}, MRLF \cite{Li&al2009}, and HA \cite{Douiri&Elbernoussi2011}. No averaged value appears in the table for HA, MDS5, and MRLF since this information is not given in \cite{Douiri&Elbernoussi2011,Helmar&Chiarandini2011,Li&al2009}. Furthermore, ``--'' marks signal that some instances were not tested by some approaches.

\renewcommand{\baselinestretch}{0.95}\large\normalsize
\begin{table}
\begin{scriptsize}
\caption{Comparisons of MASC with five state-of-the-art sum coloring algorithms} \label{table_many_algorithms}
\begin{tabular}{l@{ }r@{ }c@{ }rr@{ }c@{ }rrrrrrr}
\hline
\multicolumn{2}{c}{Graph}&&\multicolumn{2}{c}{EXSCOL \cite{Wu&Hao2012}}&&\multicolumn{2}{c}{BLS \cite{Benlic&Hao2012}}&MDS5 \cite{Helmar&Chiarandini2011}&MRLF \cite{Li&al2009}&HA \cite{Douiri&Elbernoussi2011}&\multicolumn{2}{c}{MASC} \\  
\cline{1-2}\cline{4-5}\cline{7-8}\cline{12-13}
Name & $\Sigma$&&  $\Sigma_*$& Avg. && $\Sigma_*$& Avg.&  $\Sigma_*$& $\Sigma_*$& $\Sigma_*$& $\Sigma_*$& Avg.\\
\hline
myciel3     & 21 && 21  & 21.0 &&21 & 21.0 &  21& 21& 21&21  & 21.0 \\
myciel4    &  45  &&  45  & 45.0 && 45 & 45.0 & 45 & 45 & 45 &45  & 45.0 \\
myciel5    & 93 && 93  & 93.0 &&  93 & 93.0 &  93  &93& 93& 93& 93.0 \\
myciel6     & 189 && 189  & 189.0 && 189 & 196.6 & 189  & 189& 189& 189  & 189.0 \\
myciel7    & 381  && 381  & 381.0 && 381 & 393.8 &  381& 381& 381& 381  & 381.0 \\
anna   &  276 &&  283  & 283.2 && 276 & 276.0 & 276  & 277  & --  & 276  & 276.0 \\
david    &  237 &&  237  & 238.1 && 237 & 237.0 &237& 241&--& 237  & 237.0 \\
huck   & 243 && 243  & 243.8 &&243 & 243.0 &  243& 244& 243&243  & 243.0 \\
jean   &217 &&217  & 217.3 && 217 & 217.0 & 217& 217 & --  &217  & 217.0 \\
queen5.5  & 75   && 75  & 75.0 && 75 & 75.0 &   75  &  75  &  --  & 75  & 75.0 \\
queen6.6   & 138 && 150  & 150.0 && 138 & 138.0 & 138 &  138 &  138 &138 & 138.0 \\
queen7.7    &  196 &&  196 & 196.0 &&  196 & 196.0 & 196 & 196 & -- &   196 & 196.0 \\
queen8.8    &  291  &&  291 & 291.0 &&  291 & 291.0 &  291 & 303 &-- & 291 & 291.0 \\
games120     &  443 &&  443 & 447.9 &&  443 & 443.0 &  443 & 446 & 446 & 443 &443.0 \\
miles250    &  325  &&  328 & 333.0 &&  327 & 328.8 & 325 &334 &343 & 325 & 325.0\\
miles500    &  $\le 709$ &&  709 & 714.5 && 710 & 713.3 & 712 &  715 & 755 & \textbf{705} & 705.0 \\
fpsol2.i.1    &3\,403  && -- & -- && --   & -- & 3\,403 &--&3\,405   &3\,403  & 3\,403.0  \\
mug88\_1  &178  && -- & -- && --   & -- & 178  &-- &190 &178  & 178.0    \\
mug88\_25  &178  && -- & -- && --   & -- & 178   &-- &187&178  & 178.0   \\
mug100\_1  &202  && -- & -- && --   & -- & 202  &--&211&202  & 202.0   \\
mug100\_25  & 202  & & -- & -- && --   & -- & 202  & --  & 214   & 202  & 202.0   \\
2-Insertions\_3 & 62   && -- & -- && --   & -- & 62  & --& 62 & 62   & 62.0  \\
3-Insertions\_3  &92  && -- & -- && --   & -- & 92  &-- &92 &92  & 92.0   \\
zeroin.i.2  &1\,004 && -- & -- && --   & -- & 1\,004   &-- &1\,013&1\,004  & 1\,004.0  \\
zeroin.i.3 &998  && -- & -- && --   & -- & 998  &--&1\,007 &998  & 998.0   \\
\hline
DSJC125.1   & 326  && 326 & 326.7  && 326  & 326.9 & 326& 352& -- &326  & 326.6 \\
DSJC125.5 & 1\,012   &&  1\,017  & 1\,019.7 && 1\,012 & 1\,012.9  &  1\,015 &  1\,141 &  --  & 1\,012 &  1\,020.0 \\
DSJC125.9    & 2\,503   && 2\,512 & 2\,512.0   && 2\,503  & 2\,503.0 & 2\,511  & 2\,653 & -- & 2\,503  & 2\,508.0 \\
DSJC250.1    & 973  && 985 & 985.0   && 973  & 982.5 & 977  & 1\,068 & --& 974  & 990.5 \\
DSJC250.5    &3\,219  && 3\,246 & 3\,253.9   &&3\,219  & 3\,248.5 & 3\,281&3\,658 & -- &  3\,230 & 3\,253.7   \\
DSJC250.9   & $\le 8\,286$  && 8\,286 & 8\,288.8  &&  8\,290  & 8\,316.0 & 8\,412&8\,942&-- &\textbf{8\,280} & 8\,322.7   \\

DSJC500.1    & 2\,850  && 2\,850 &  2\,857.4   &&  2\,882  & 2\,942.9 & 2\,951  & 3\,229 & --& 2\,940  & 3\,013.4 \\
DSJC500.5    &10\,910  &&10\,910 & 10\,918.2   &&11\,187  & 11\,326.3 & 11\,717 &12\,717 & -- &  11\,101 &  11\,303.5  \\
DSJC500.9   & 29\,912  &&29\,912 & 29\,936.2  &&  30\,097  &  30\,259.2 &30\,872 &32\,703 &-- &29\,994 & 30\,059.1  \\

flat300\_20\_0     & 3\,150  && 3\,150 & 3\,150.0 && -- & -- & -- & -- & -- & 3\,150 & 3\,150.0 \\
flat300\_26\_0    & 3\,966 && 3\,966 & 3\,966.0 && -- & -- & -- & -- & -- & 3\,966 & 3\,966.0 \\
flat300\_28\_0     & $\le 4\,282$  && 4\,282 & 4\,286.1 &&  -- & -- & -- & -- & -- & \textbf{4\,238} & 4\,313.4 \\
le450\_15a      & 2\,632  && 2\,632 & 2\,641.9 && -- & -- & -- & -- & -- &2\,706  & 2\,742.6   \\
le450\_15b      & 2\,642 && 2\,642 & 2\,643.4 && -- & -- & --  & -- & --  & 2\,724 & 2\,756.2  \\
le450\_15c      & $\le 3\,866$ && 3\,866 & 3\,868.9 && -- & -- & --  & -- & --  & \textbf{3\,491} & 3\,491.0  \\
le450\_15d     & $\le 3\,921$   && 3\,921 & 3\,928.5 && -- & -- & --   & -- & -- & \textbf{3\,506} & 3\,511.8  \\
le450\_25a      & 3\,153 && 3\,153 & 3\,159.4 && -- & -- & --   & -- & -- & 3\,166 & 3\,176.8  \\
le450\_25b       & 3\,366  && 3\,366 & 3\,371.9 && -- & -- & --   & -- & -- & 3\,366 & 3\,375.1  \\
le450\_25c      & 4\,515  && 4\,515 & 4\,525.4 && -- & -- & --   & -- & -- & 4\,700 & 4\,773.3  \\
le450\_25d    & 4\,544  && 4\,544 & 4\,550.0 && -- & -- & --   & -- & -- & 4\,722 & 4\,805.7  \\
\hline
\end{tabular}
\end{scriptsize}
\end{table}
\renewcommand{\baselinestretch}{1.0}\large\normalsize

\begin{table}
\begin{scriptsize}
\caption{MASC vs. five state-of-the-art sum coloring algorithms}
\label{comparisons_other_algorithms}
\begin{tabular}{lcrrr}
\hline
\multirow{2}{*}{Competitor} & \multirow{2}{*}{$\#G$} & \multicolumn{3}{c}{Results of MASC ($\Sigma_*$)}\\
\cline{3-5}
&& Better   & Equal   & Worse	\\
\hline
EXSCOL \cite{Wu&Hao2012} & 36            & 12             & 16       & 8 \\
BLS \cite{Benlic&Hao2012}   & 25            & 5             & 17         & 3 \\
MDS5 \cite{Helmar&Chiarandini2011}  & 34            &9             & 25       & 0 \\
MRLF  \cite{Li&al2009} & 25            & 16          & 9         & 0 \\
HA   \cite{Douiri&Elbernoussi2011} & 19            & 10            & 9        & 0 \\
\hline
\end{tabular}
\end{scriptsize}
\end{table}

Since the reference algorithms give only results for a (small) subset of the considered benchmark, it is difficult to analyze the performance of these algorithms by statistical tests. Hence, we compare the performance between MASC and these reference algorithms one by one and summarize the comparisons in Table~\ref{comparisons_other_algorithms}. The first column of Table \ref{comparisons_other_algorithms} indicates the name of the reference heuristics, followed by the number $\#G$ of graphs tested by each algorithm and shown in Table \ref{table_many_algorithms}. The last three columns give the number of times MASC reports a better, equal, or worse result compared to each reference algorithm.

From Table \ref{comparisons_other_algorithms}, it can be observed  that MASC obtains absolutely no worse results than MDS5, MRLF, and HA (see the last three lines). Furthermore, MASC gets better results than these algorithms for 9, 16, and 10 instances respectively. Our algorithm is also quite competitive with EXSCOL and BLS which are the most recent and effective methods since it obtains better or equivalent results for 28 and 22 graphs respectively. MASC reaches worse results than EXSCOL and BLS only for 8 and 3 graphs respectively.

\subsection{Experiment on Large Graphs}
\label{subsec_experiments_large}

We turn now our attention to the performance of our MASC algorithm to color large graphs with at least 500 vertices. These large graphs are known to be quite difficult for almost all the existing sum coloring approaches except EXSCOL which dominates the other heuristics particularly on large graphs. We show a new experiment with MACS applied to color 17 large graphs. In this experiment, we run MASC 10 times on each graph under exactly the same condition as in Section \ref{subsec_computational_results}. The only difference is that we use the solution of EXSCOL\footnote{Available at \url{http://www.info.univ-angers.fr/pub/hao/exscol.html}} as one of MASC's 10 initial solutions while the 9 other initial solutions are generated according to the procedure described in Section \ref{subsec_sol_initial_population}. With this experiment, we aimed to investigate two interesting questions. Is it possible for MASC to improve the results of the powerful EXSCOL algorithm? Does the initial population influence the performance of MASC? The computational outcomes of this experiment are provided in Table \ref{table_MASC_large}.

\begin{table}
\begin{scriptsize}
\caption{Results of MASC on 17 large graphs with at least 500 vertices}
\label{table_MASC_large}
\begin{tabular}{lrrrcrrcrrrrr}
\hline
\multicolumn{4}{c}{Characteristics of the graphs} && \multicolumn{2}{c}{EXSCOL} && \multicolumn{5}{c}{MASC}\\
\cline{1-4}\cline{6-7}\cline{9-13}
Name & $|V|$ & $|E|$ & $\Sigma$  && $\Sigma_*$ & Avg.&& $k$ &$\Sigma_*(k_*)$ & Avg. & $\sigma$ & $t$ \\
\hline
DSJC500.1    & 500 & 12\,458   &2\,850 &&2\,850 & 2\,857.4&&12&  \textbf{2\,841(14)} & 2\,844.1&3.2  & 28.9  \\
DSJC500.5    & 500 & 62\,624  &10\,910 &&10\,910 & 10\,918.2  &&48& \textbf{10\,897(51)}  & 10\,905.8 &4.6 & 73.3 \\
DSJC500.9    & 500 & 112\,437  &29\,912 &&29\,912 &29\,936.2 &&126&  \textbf{ 28\,896(131)} & 29\,907.8&5.8  & 59.0  \\
DSJC1000.1    &1\,000 & 49\,629   & 9\,003 && 9\,003 &9\,017.9 &&20&  \textbf{8\,995(22)}  & 9\,000.5&3.0  & 70.7  \\
DSJC1000.5    &1\,000 & 249\,826  &37\,598 &&37\,598 & 37\,673.8  &&83&  \textbf{37\,594(87)} &37\,597.6  &1.2 & 200.4 \\
DSJC1000.9    &1\,000 & 449\,449  &103\,464 &&103\,464 &103\,531.0&&223& 103\,464(231) & 103\,464.0& 0.0 & 125.9  \\

flat1000\_50\_0    & 1\,000 & 245\,000 & 25\,500 && 25\,500 &25\,500.0 &&50& 25\,500(50) & 25\,500.0&0.0 & 0.1 \\
flat1000\_60\_0    & 1\,000 & 245\,830 & 30\,100 && 30\,100 &30\,100.0 &&60& 30\,100(60) & 30\,100.0&0.0 & 114.6 \\
flat1000\_76\_0    & 1\,000 & 246\,708 & 37\,167 && 37\,167 &37\,213.2 &&82& 37\,167 (85) &37\,167.0 &0.0  &1.1  \\

latin\_sqr\_10    & 900 &307\,350 &42\,223 &&42\,223 &42\,392.7 &&98&  \textbf{41\,444(100)}& 41\,481.5&19.1  & 101.2 \\

wap05    & 905 & 43\,081 &13\,680 &&13\,680 &13\,718.4 &&50&  \textbf{13\,669(51)} & 13\,677.8&3.7  & 3.3 \\
wap06    & 947 & 43\,571 &13\,778 &&13\,778 &13\,830.9 &&46&  \textbf{13\,776(48)} & 13\,777.8&0.6 & 4.1 \\
wap07    & 1\,809 & 103\,368 &28\,629 &&28\,629 &28\,663.8 &&46&  \textbf{28\,617(50)} & 28\,624.7&3.8 & 12.4\\
wap08    & 1\,870 & 104\,176 &28\,896 &&28\,896 &28\,946.0 &&45&  \textbf{28\,885(50)} & 28\,890.9&3.2 & 15.1 \\

qg.order30    & 900 & 26\,100 &13\,950 &&13\,950 &13\,950.0 &&30&  \textbf{12\,581(31)} & 12\,641.3&45.7 & 4.2\\
qg.order40    &1\,600 &62\,400 &32\,800 &&32\,800 &32\,800.0 &&40&  32\,800(40)& 32\,800.0&0.0 & 11.8 \\
qg.order60    & 3\,600 & 212\,400 &110\,925 &&110\,925 &110\,993.0 &&60&  \textbf{109\,800(60)} &109\,800.0 &0.0 &290.6  \\

\hline
\end{tabular}
\end{scriptsize}
\end{table}

In Table \ref{table_MASC_large}, column 4 presents the best known result ($\Sigma$) in the literature, columns 5--6 present the best result ($\Sigma_*$) and the average coloring sum (Avg.) of EXSCOL and columns 7--11 present detailed computational results of our MASC algorithm: Best result obtained ($\Sigma_*$) with the number of required colors ($k_*$), average coloring sum (Avg.), standard deviation ($\sigma$), and average running time to reach $\Sigma_*$ ($t$, in minutes). One notices that the values of columns 4 and 5 are identical. This is because EXSCOL is the single approach in the literature able to attain these results. 


Table \ref{table_MASC_large} shows that with the help of its search mechanism, our MASC algorithm is able to further improve the best known results of 12 instances (entries in bold). This is remarkable given that no previous approach can even equal these results. Moreover, if we contrast the results of the three DSJC500.$x$ graphs ($x=1,5,9$)  reported in Tables \ref{table_MASC} and \ref{table_MASC_large}, it is clear that the initial population impacts directly MASC's outcomes. This indicates that the performance of MASC could be further improved by using a more powerful coloring algorithm to generate the initial solutions of its population.

\section{Analysis of MASC}
\label{Sec_analysis_MASC}

In this section, we investigate the influence of two important ingredients of the proposed memetic algorithm, i.e., the multi-parent crossover operator and the combined neighborhood. Experiments were based on 16 selected graphs of different types, for which some reference algorithms cannot achieve the best known results. Hence, these selected instances can be considered to be difficult and representative.

\subsection{Influence of the Multi-parent Crossover Operator}
\label{subsec_sol_influence_Crossover}

For our memetic algorithm, it is relevant to evaluate the effectiveness of its crossover operator. To verify this, we carry out experiments on the 16 selected graphs and run both MASC (using the MGPX crossover) and DNTS (without MGPX) for 30 times (with the same parameter $p_1$, $p_2$ and $p_3$ settings as defined in Table \ref{Parameter_Settings}). The DNTS (without MGPX) starts with a single solution which is generated for MASC. DNTS stops when a maximum number of $5 \times 10^5$ iterations in order to make sure that MASC and DNTS are given the same search effort. The results are given in Table \ref{table_MASC_TS}. 

From Table \ref{table_MASC_TS}, one notices that DNTS equals and improves respectively 5 and 3 best known results while MASC equals and improves respectively 5 and 11 best known results. Furthermore, the last column \emph{tt} (t-test) indicates whether the observed difference between MASC and DNTS is statistically significant when a 95\% confidence t-test is performed in terms of the best result obtained ($\Sigma_*$). The t-test indicates that MASC is statistically better than DNTS for 12 out of 16 cases except for the instances where DNTS can achieve the best known results ($\Sigma$). These comparative results provide clear evidences that the MGPX crossover operator plays an important role in the MASC algorithm.

\begin{table}
\begin{scriptsize}
\caption{Comparative results of MASC and DNTS} \label{table_MASC_TS}
\begin{tabular}{lrcrrrrrcl}
\hline
\multicolumn{2}{c}{Graph} && \multicolumn{2}{c}{MASC} & &\multicolumn{2}{c}{DNTS} &\multirow{2}{*}{\emph{tt}}\\  
\cline{1-2}\cline{4-5} \cline{7-8} 
Name&$\Sigma$&& $\Sigma_*$ & Avg. & & $\Sigma_*$ & Avg. \\
\hline
anna              & 276 && \emph{276} & 276.0 & &\emph{276}   & 276.0 & N\\
queen6.6          & 138 && \textit{138} & 138.0 && \emph{138}  & 138.0 &N\\
miles250          & 325    && \textit{325} & 325.0 & &\emph{325} & 325.0 &N \\
miles500          & $\le 709$   && \textbf{705} & 705.0 & & \textbf{705} & 705.6 &Y\\
DSJC125.1         & 326  && \textit{326}  & 326.6 & & \emph{326}  & 328.6 &Y\\
DSJC125.5         & 1\,012   && \textbf{1\,012}  & 1\,020.0  & & 1\,016  & 1\,029.8  &Y\\
DSJC125.9         & 2\,503 && \textbf{2\,503}  & 2\,508.0   & & 2\,506  & 2\,530.1  &Y \\
DSJC250.1         & 973  && \textbf{974}  & 990.5    &  & 981  & 997.7   &Y \\
DSJC250.5          &3\,219  &&\textbf{3\,230}  & 3\,253.7  & &3\,234  & 3\,301.7  &Y\\
DSJC250.9         & $\le 8\,286$  && \textbf{8\,280}  & 8\,322.7   && 8\,321  & 8\,381.9   &Y\\
flat300\_26\_0    & 3\,966 && \textit{3\,966} & 3\,966.0  & & \emph{3\,966} & 3\,966.0  &N\\
flat300\_28\_0    &  $\le 4\,282$ &&  \textbf{4\,238} & 4\,313.4 & &  4\,303 & 4\,406.3 &Y\\
le450\_15c        & $\le 3\,866$  && \textbf{3\,491} & 3\,491.0   & & \textbf{3\,491} & 3\,492.1  &Y \\
le450\_15d        &  $\le 3\,921$ &&  \textbf{3\,506} & 3\,511.8  &  &  \textbf{3\,506} & 3\,515.0 &Y \\
le450\_25c       & 4\,515  && \textbf{4\,700} & 4\,773.3  & & 4\,749 & 4\,803.9  &Y\\
le450\_25d        & 4\,544 && \textbf{4\,722} & 4\,805.7 & & 4\,784 & 4\,835.3 &Y\\
\hline
\end{tabular}
\end{scriptsize}
\end{table}

\subsection{Influence of the Neighborhood Combination}
\label{subsec_sol_influence_neighborhood_combination}

The neighborhood is an important element that influences the local search procedure. Our proposed algorithm relies on two different neighborhoods: $N_1$ (neighborhood based on connected components) and $N_2$ (neighborhood based on one-vertex-move) which are explored in a token-ring way (see Section \ref{subsec_sol_TS}). In this section, we investigate the interest of this combined use of the two neighborhoods. For this purpose, we carried out experiments on the 16 selected graphs to compare the original Double-Neighborhood Tabu Search (DNTS) with two variants which uses only one neighborhood $N_1$ or $N_2$. We use below TS$_{N1}$ and TS$_{N2}$ to denote these two variants. These three TS procedures (DNTS, TS$_{N1}$ and TS$_{N2}$) are run under the same stop condition, i.e. limited to $5 \times 10^5 $ iterations.

We run 30 times these TS procedures to solve each of the 16 selected graphs and report the computational outcomes (the best and average results) in Table \ref{table_Multi_Single}. One easily observes that DNTS obtains better or equal results compared to TS$_{N1}$ and TS$_{N2}$ for all the instances in terms of the best known result ($\Sigma_*$) and the average result (Avg.). The t-test \emph{$tt_i$} ($i = 1, 2$) in the last two columns confirms that with a 95\% confidence level DNTS is slightly or significantly better than  TS$_{N1}$ and TS$_{N2}$. This experiment demonstrates thus the advantage of the token-ring combination of the two neighborhoods compared to each individual neighborhood.

\begin{table}[h]
\begin{scriptsize}
\caption{Comparative results of the LS improvement method according to the neighborhood employed} \label{table_Multi_Single}
\begin{tabular}{lrrcrrcrrcrcr}
\hline
\multirow{2}{*}{Graph} & \multicolumn{2}{c}{DNTS} & &\multicolumn{2}{c}{TS$_{N2}$} & &\multicolumn{2}{c}{TS$_{N1}$} &\multirow{2}{*}{\emph{$tt_2$}} &\multirow{2}{*}{\emph{$tt_1$}}\\  
\cline{2-3} \cline{5-6} \cline{8-9} 
& $\Sigma_*$& Avg. & & $\Sigma_*$& Avg. & & $\Sigma_*$& Avg. \\
\hline
anna              &276 &276.0 &  &282 & 285.8 &  &276   & 276.0  &Y & N \\
queen6.6      & 138 & 138.0 &  &138 & 138.4 &   & 138  & 138.0  &Y &N\\
miles250        & \textbf{325} & 325.0 &   &346 & 361.6  &   & 335 & 340.7 &Y &Y\\
miles500        & \textbf{705} & 705.6 &  &722 & 736.0 &  & 719 & 730.9 &Y &Y \\
DSJC125.1        & \textbf{326}  & 328.6   & & 334  & 340.8   & & 329  & 334.0 &Y &Y\\
DSJC125.5        & \textbf{1016}  & 1029.8   & & 1\,031  & 1\,045.1   & & 1\,020  & 1\,031.8 &Y &N\\
DSJC125.9        & \textbf{2\,506}  & 2\,530.1   & & 2\,514  & 2\,557.6   & & 2\,512  & 2\,538.3 &Y &N\\
DSJC250.1         & \textbf{981}  & 997.7    &   & 1\,004  & 1\,021.3    &   & 1\,022  & 1\,039.9  &Y &Y \\
DSJC250.5          &\textbf{3\,234}  & 3\,301.7  &  &3\,271  & 3\,323.9  &  &3\,260  & 3\,306.5 &Y &N \\
DSJC250.9          &\textbf{8\,321}  & 8\,381.9  &  &8\,347  & 8\,405.6  &  &8\,318  & 8\,387.5 &Y &N \\
flat300\_26\_0      & 3\,966 & 3\,966.0  & & 3\,966 & 3\,966.0 & & 3\,966 & 3\,966.0 &N &N \\
flat300\_28\_0      &  \textbf{4\,303} & 4\,406.3 & &  4\,347 & 4\,427.8 & &  4\,332 & 4\,435.5  &N &Y\\
le450\_15c          & \textbf{3\,491} & 3\,492.5   & & 3\,503  & 3\,517.2  & & 3\,508 & 3\,551.8    &Y &Y\\
le450\_15d           &  \textbf{3\,506} & 3\,515.0  & &  3\,528 & 3\,538.5 & &  3\,526  & 3\,568.2 &Y &Y \\
le450\_25c        &\textbf{4\,749} & 4\,803.9 &   &4\,828 & 4\,893.9 & &5\,005 & 5\,067.4  &Y &Y\\
le450\_25d       & \textbf{4\,784} & 4\,835.3  &  & 4\,848 & 4\,907.0  &  & 5\,035 & 5\,119.1  &Y &Y\\
\hline
\end{tabular}
\end{scriptsize}
\end{table}

\section{Conclusion}
\label{Sec_Conclusion}

In this paper, we presented a memetic algorithm (MASC) to deal with the minimum sum coloring problem (MSCP). The proposed algorithm employs an effective tabu search procedure with a combination of two neighborhoods, a multi-parent crossover operator and a population updating mechanism to balance intensification and diversification.

We assessed the performance of MASC on 77 well-known graphs from the DIMACS and COLOR 2002-2004competitions. MASC can improve 17 best known upper bounds including 11 large and very hard graphs with at least 500 vertices while equaling 30 previous best results. We also report upper bounds for the 18 remaining graphs for the first time. Compared with five recent and effective algorithms which cover the best known results for the tested instances, our MASC algorithm remains quite competitive.

Furthermore, we investigated two important components of the proposed algorithm. The experiments demonstrate the relevance of the multi-parent crossover operator and the combined neighborhood for the overall performance of MASC. 


\section*{Acknowledgment}
The work is partially supported by the RaDaPop (2009-2013) and LigeRo projects (2009-2013) from the Region of Pays de la Loire (France). Support for Yan Jin from the China Scholarship Council is also acknowledged.

\end{document}